\documentclass[10pt,superscriptaddress,prl,aps,twocolumn,showpacs,amsmath,amssymb]{revtex4-1}
\usepackage{graphicx}
\usepackage[caption=false]{subfig}
\usepackage{hyperref}
\usepackage[capitalise]{cleveref}
\usepackage{dcolumn}
\usepackage{bm}

\newcommand{\pavia}[1]{\affiliation{Dipartimento di Fisica, Università degli Studi di Pavia, I-27100 Pavia, Italy}}
\newcommand{\infn}[1]{\affiliation{INFN Sezione di Pavia, I-27100 Pavia, Italy}}
\newcommand{\mainz}[1]{\affiliation{Institut f\"ur Kernphysik, Johannes Gutenberg-Universit\"at Mainz, D-55099 Mainz,Germany}}
\newcommand{\regensburg}[1]{\affiliation{Institut f\"ur Theoretische Physik, Universit\"at Regensburg, D-93040 Regensburg, Germany}}
\newcommand{\aE}{\alpha_{E1}}
\newcommand{\bM}{\beta_{M1}}

\begin{document}

\title{First concurrent extraction of the leading-order scalar and spin proton polarizabilities}%

\author{E.~Mornacchi}\email{e.mornacchi@uni-mainz.de}\mainz \\
\author{S.~Rodini}  \regensburg \\
\author{B.~Pasquini} \pavia\\ \infn\\
\author{P.~Pedroni} \infn

\date{\today}

\begin{abstract}
We performed the first simultaneous extraction of the six leading-order proton polarizabilities. We reached this milestone thanks to both new high-quality
experimental data and an innovative bootstrap-base fitting method. These new results provide a self-consistent and fundamental benchmark for all future theoretical and experimental polarizability estimates.
\end{abstract}

\maketitle


\section{\label{sec:Intro} Introduction}
Understanding the hadron structure in the non-perturbative regime of quantum chromodynamics (QCD)
is one of the major challenge of modern physics. 
We can classify hadrons in terms of their global properties, such as mass and spin,
but we can not fully explain how these properties emerge from the underlying dynamics
of the hadron's interior. A clean probe to investigate the internal structure of hadrons
is the Compton scattering process that gives access to observables with a clear interpretation
in terms of structure-dependent properties of the hadrons.
In particular, real Compton scattering (RCS) at low energies is parametrized
by polarizabilities that describe the response of the charge and magnetization
distributions inside the nucleon to an applied quasi-static electromagnetic field. 
These structure constants are fundamental properties of the nucleon and their
determination has driven a relevant experimental effort in the last few
years~\cite{A2:2014iky,A2:2016nio,A2:2019bqm,A2:2021rcs,higs2022}.

The effective multipole interactions for the coupling of the electric ($\vec E$)
and magnetic ($\vec{H}$) fields of the photon with the internal structure of the
nucleon is described at leading order in terms of the electric ($\aE$) and magnetic
($\bM$) scalar polarizabilities~\cite{Babusci:1998ww,Holstein:1999uu}:
\begin{equation}
	H^{(2)}_{eff} = -4\pi \Bigl[\frac{1}{2} \alpha_{E1} \vec{E}^2 + \frac{1}{2} \beta_{M1} \vec{H}^2 \Bigr],
	\label{eq:H2}
\end{equation}
while the four spin polarizabilities ($\gamma_{\dots}$) show up in the subleading terms:
\begin{align*}
	H^{(3)}_{eff} &= -4\pi \Bigl[ \frac{1}{2} \gamma_{E1E1} \vec{\sigma} \cdot (\vec{E} \times \dot{\vec{E}}) + \frac{1}{2} \gamma_{M1M1} \vec{\sigma} \cdot (\vec{H} \times \dot{\vec{H}}) \\
	&\quad- \gamma_{M1E2} E_{ij} \sigma_{i}H_{j} + \gamma_{E1M2} H_{ij} \sigma_{i}E_{j} \Bigr],
	\stepcounter{equation}\tag{\theequation}\label{eq:H3}
\end{align*}
where $\vec{\sigma}$ are the proton’s Pauli spin matrices, $\dot{\vec{E}} = \partial_{t} \vec{E}$ and $E_{ij} = \frac{1}{2}(\nabla_i E_j + \nabla_j E_i$) are partial derivatives with respect to time and space, respectively.

In addition to being fundamental properties of the nucleon, polarizabilities play
a profound role in precision atomic physics, in the evaluation of the nuclear 
corrections to atomic energy levels~\cite{Drell:1966kk, Bernabeu:1982qy, Faustov:1999ga, Martynenko:2005rc, Carlson:2011zd}
and in astrophysics, influencing neutron star properties~\cite{Bernabeu:1974zu}.

Despite their evident importance in a broad range of physics topics, up to now a
self-consistent experimental extraction of all the different polarizability values
has not been possible, due to the scarce quality of the available database
(see, for instance, Ref.~\cite{Pasquini:2019nnx}).
In all existing fits of the RCS data, some of the polarizabilities have been fixed either
using theoretical calculations~\cite{Griesshammer:2012we,McGovern:2012ew,Lensky:2014efa,Griesshammer:2016}
or empirical evaluations from other reactions~\cite{Krupina:2018}, or, at most,
have been constrained to vary within certain intervals~\cite{Pasquini:2019nnx}.
The situation has recently improved with the first measurements of the
double-polarization observables $\Sigma_{2x}$~\cite{A2:2014iky} and
$\Sigma_{2z}$~\cite{A2:2019bqm} and new data for the unpolarized differential
cross sections and the single-polarized $\Sigma_{3}$
asymmetry~\cite{A2:2016nio,A2:2021rcs,higs2022}. The beam asymmetry is defined as~\citep{Babusci:1998ww}:
\begin{equation}
\Sigma_{3} = \dfrac{d\sigma_{\parallel} - d\sigma_{\perp}} {d\sigma_{\parallel} + d\sigma_{\perp}},
\label{S3}
\end{equation}
where $d\sigma_{\parallel(\perp)}$ is the polarized cross section obtained with a photon beam polarized parallel (or perpendicularly) to the scattering plane and an unpolarized target. In a similar way, the double-polarization asymmetries can be defined as:
\begin{equation}
	\Sigma_{2x} = \dfrac{d\sigma^R_{+x} - d\sigma^L_{\pm x}}{d\sigma^R_{+x} + d\sigma^L_{\pm x}},~~\text{and}~~\Sigma_{2z} = \dfrac{d\sigma^R_{+z} - d\sigma^L_{\pm z}}{d\sigma^R_{+z} + d\sigma^L_{\pm z}},
	\label{eq:S2}
\end{equation}
where $d\sigma^{R(L)}_{\pm x}$ is the polarized cross section obtained with circular right-handed (left-handed) photon polarization and target spin aligned transversely $(\pm x)$ with respect to the incident beam direction, while $d\sigma^{R(L)}_{\pm z}$ is obtained with the target spin aligned longitudinally $(\pm z)$ with respect to the incident beam direction.

In particular, the work from Mornacchi~\textit{et al.}~\cite{A2:2021rcs}
provides the highest statistics proton RCS single dataset ever obtained
below the pion photoproduction threshold, with 60 unpolarized differential
cross section points and 36 beam asymmetry points measured over a large angular range and
with small statistical and systematic errors.
Therefore, it represents a significant improvement for a more accurate 
extractions of all the different polarizability values.

In this paper we present the first consistent and simultaneous fit 
of the six leading-order static proton polarizabilities. It has been obtained
thanks to both the new experimental data and an innovative 
bootstrap-base fitting method~\cite{Pedroni:2019dlg}.
This algorithm has already been deployed successfully for the extraction of the proton scalar
dynamical and static polarizabilities from low-energy RCS data~\cite{Pasquini:2018,Pasquini:2019nnx}.
The theoretical framework used for this extraction is based on fixed-$t$ subtracted
dispersion relations (DRs)~\cite{Drechsel:1999rf,Holstein:1999uu,Pasquini:2007hf}.
The theoretical uncertainties associated to the model dependence of our results are
also evaluated by using, as input, pion photoproduction amplitudes obtained from
three different partial wave analyses
(PWAs) of the available experimental data: BnGa-2019~\cite{Anisovich:2016vzt},
MAID-2021~\cite{Drechsel:2007if, MAID21}, and SAID-MA19~\cite{A2:SAID}.
This is the very first time that such a comprehensive and self-consistent study on the 
simultaneous extraction of all the six leading-order proton polarizabilities from
RCS data is performed.

\section{\label{sec:fit} Database selection and fit procedure}
The proton RCS database used for this work consists
of two main sets: the unpolarized differential cross section data,
and the (single and double) polarization asymmetries.
The former can be further divided into low- and high-energy data,
namely data for which the incoming photon energy $E_{\gamma}$ is below or
above the pion photoproduction threshold ($\sim 150$~MeV), respectively.

For the low-energy set, in addition to the new data from Refs.~\cite{A2:2021rcs,higs2022},
we used the same selection extensively discussed in a previous work by
Pasquini~\textit{et al.}~\cite{Pasquini:2019nnx}, which includes datasets from 
Refs.~\cite{Oxley:1958zz,Hyman:1959zz,GOLDANSKY1960473,Bernardini:1960wya,Pugh:1957zz,Baranov:1974ec,Baranov:1975ju,Federspiel:1991yd,ZIEGER199234,Hallin:1993ft,MacGibbon:1995in,deLeon:2001dnx}.
For the high-energy set, thanks to the DR model used for the theoretical
framework, we were able to consider data measured up to
$E_{\gamma} = 300$~MeV (corresponding to the
$2\pi$ photoproduction threshold), thus extending the range used in the fits of
Refs.~\cite{McGovern:2012ew,Pasquini:2019nnx}.
Furthermore, only for these high-energy data, we decided to narrow the selection to the new-generation experiments,
namely the measurements performed using tagged photon facilities
(see, for instance, the review of Ref.~\cite{Schumacher:2005an}).
Their main advantage, in addition to a more reliable photon flux determination,
is that the incoming photon energy is known with a resolution of a few MeV.
This is an essential ingredient to reject the overwhelming background coming from the
single $\pi^0$ photoproduction channel that has a cross section by two orders of magnitude
higher than Compton scattering process, and can mimic the Compton signature
when one of the two photons coming from the $\pi^0$ decay escapes the particle detection.
The available datasets for the unpolarized cross section come from two different facilities: 
MAMI~\cite{Peise:1996zz,Molinari:1996zw,Wissmann:1999vi,Wolf:2001ha,Galler:2001ht,Camen:2001st}
(with also few data above threshold from Ref.~\cite{deLeon:2001dnx})
and LEGS~\cite{Tonnison:1998mi,Blanpied:2001ae}.

The polarization observables have enhanced sensitivity to the spin
polarizabilities~\cite{Pasquini:2007hf,Griesshammer:2017txw}, hence are crucial for the
extraction of these structure constants.
The adopted datasets include three different polarization observables:
$\Sigma_{2x}$~\cite{A2:2014iky}, $\Sigma_{2z}$~\cite{A2:2019bqm}, and $\Sigma_3$ both
below~\cite{A2:2016nio,A2:2021rcs,higs2022} and above~\cite{Blanpied:2001ae}
pion photoproduction threshold.

The fit to extract the leading-order scalar and spin polarizabilities was performed using
a bootstrap-based method~\cite{Pedroni:2019dlg} that consists of randomly generating 
$N$ Monte Carlo replicas of the fitted experimental database, where each data point
$e_{i,j}^{(0)}$ is replaced by:
\begin{equation}
   e_{i,j}^{(0)} \rightarrow e_{i,j}^{(b)} = (1 + \delta_{j,b})(e_{i,j}^{(0)} + r_{i,j,b} \sigma^{(0)}_{i,j}).
\end{equation}
The indices $i$, $j$, and $b$ run over the number of data points in each dataset,
the number of datasets, and the bootstrap replica, respectively;
$r_{i,j,b}$ is a random number extracted from the normal distribution $\mathcal{N}(0,1)$,
and $\delta_{j,b}$ is a random variable that accounts for the effect of the
common systematic errors, independently for each dataset.
From each of these simulated databases, a set of fitted parameters is extracted.
The mean and standard deviation of the obtained distributions give then the best value
and the error for each of the fitted parameters.
This technique offers several advantages compared to other fitting procedures,
especially when different datasets are used together, as in the present work: 
i) a straightforward inclusion of common systematic uncertainties without any \textit{a priori} assumption on their distributions and without introducing any additional fit parameter;
ii) the probability distribution of the fit parameters (often non-gaussian) is obtained directly by the procedure itself;
iii) the uncertainties on possible nuisance model parameters are easily and directly taken into account in the sampling procedure;
iv) the correct fit $p$-value is always provided when systematic uncertainties are present and in all the other cases when the goodness-of-fit distribution is not given by the $\chi^2$-distribution.

In the first step of our analysis, we checked the consistency of the selected database
by looking at the distribution of the normalized residual for each of the largest datasets
(e.g., with more than 40 data points): the unpolarized cross section from 
the A2~\cite{A2:2021rcs}, the TAPS~\cite{deLeon:2001dnx}, and the LARA~\cite{Wolf:2001ha}
Collaborations; and the unpolarised cross section and beam asymmetry from the LEGS~\cite{Blanpied:2001ae} Collaboration.

Since the cross section data from LARA and LEGS are known to be in 
significant disagreement between each other (see, for instance, Ref.~\cite{Griesshammer:2012we}),
we performed a preliminary test by alternatively including LARA or LEGS data in the fit database and simultaneously fitting 
all six polarizabilities using the MAID-2021~\cite{Drechsel:2007if,MAID21} multipole solution.

\begin{figure}[t]%
    \begin{center}%
        \includegraphics[width=0.48\textwidth,angle=0]{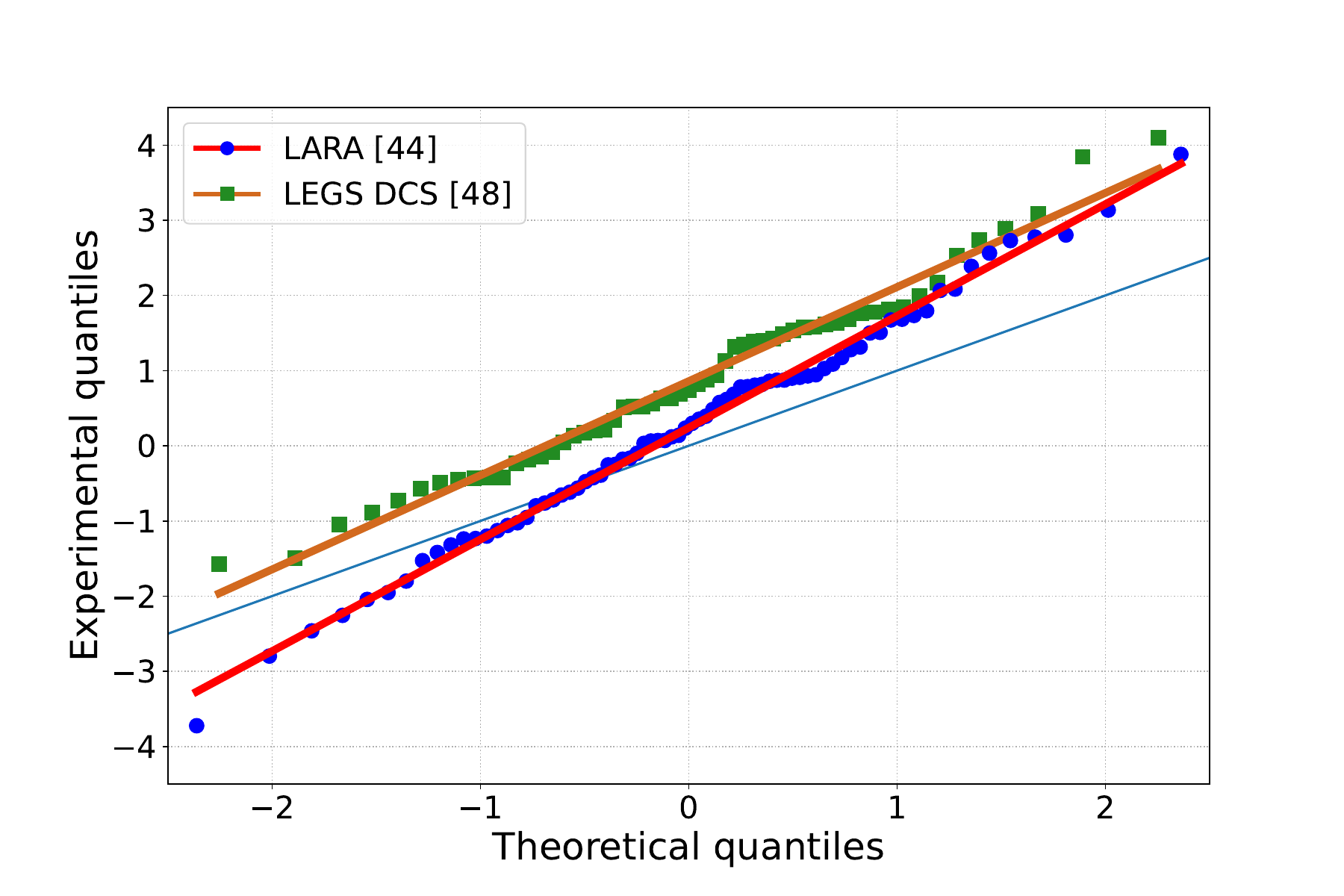}%
    \end{center}%
    \caption{Probability plots using the normalized residuals for the unpolarized cross section from the LARA~\cite{Wolf:2001ha} and the LEGS~\cite{Blanpied:2001ae} Collaborations, obtained from a global fit using the MAID-2021 multipole solution~\cite{Drechsel:2007if,MAID21}. The points are expected to lie on the bisector (in cyan) if their residuals are normally distributed. The red and orange lines show a linear regression fit to the points from the LARA and LEGS data, respectively, to help the comparison with the expected distribution.}
    \label{fig:probplot}
\end{figure}

For each of the two configurations, 
we took the output polarizability best-values, calculated the
residual distribution for each dataset, and produced a
probability plot~\cite{chambers:1983} for assessing the residual normal distribution.
In all cases, the residuals were found to follow fairly well the expected normal distribution for all the 
selected datasets (see Fig.1 of the Supplemental Material~\cite{suppMat}, where the probability plots refer 
to the test without the LEGS data), except for the unpolarized cross section from
both the LEGS and LARA Collaborations, as shown in \cref{fig:probplot}.
We repeated the same test using the SAID-MA19~\cite{A2:SAID} and BnGa-2019~\cite{Anisovich:2016vzt} PWAs 
and obtained very similar results.
Given also the small fit $p$-values, i.e., $\simeq 5\cdot 10^{-4}$ and $\simeq 1\cdot 10^{-2}$ for the LEGS and LARA data, respectively, with all the PWA inputs, we excluded both sets from the 
database of the present work.

Although the data for the photon asymmetry $\Sigma_3$ above threshold are from
the same dataset of the LEGS unpolarized cross section, they give a consistent residual plot 
(see Fig.~1 of the Supplemental Material~\cite{suppMat}). 
Such an agreement is not surprising since, 
as noted, e.g.\@, in Ref.~\cite{Griesshammer:2012we}, most of the possible systematic biases cancel out
in an asymmetry observable.

Taking all these considerations together, 25 datasets for a total of 388 points 
were included in the fit. The angular and energy coverage of all the used datasets
are listed in the Supplemental Material~\cite{suppMat}.

\section{\label{sec:Results} Results and Discussion}
\begin{figure*}[ht]%
    \begin{center}%
    \subfloat[\label{fig:Results:alpha}$\alpha_{E1}$]{\includegraphics[width = 0.33\linewidth]{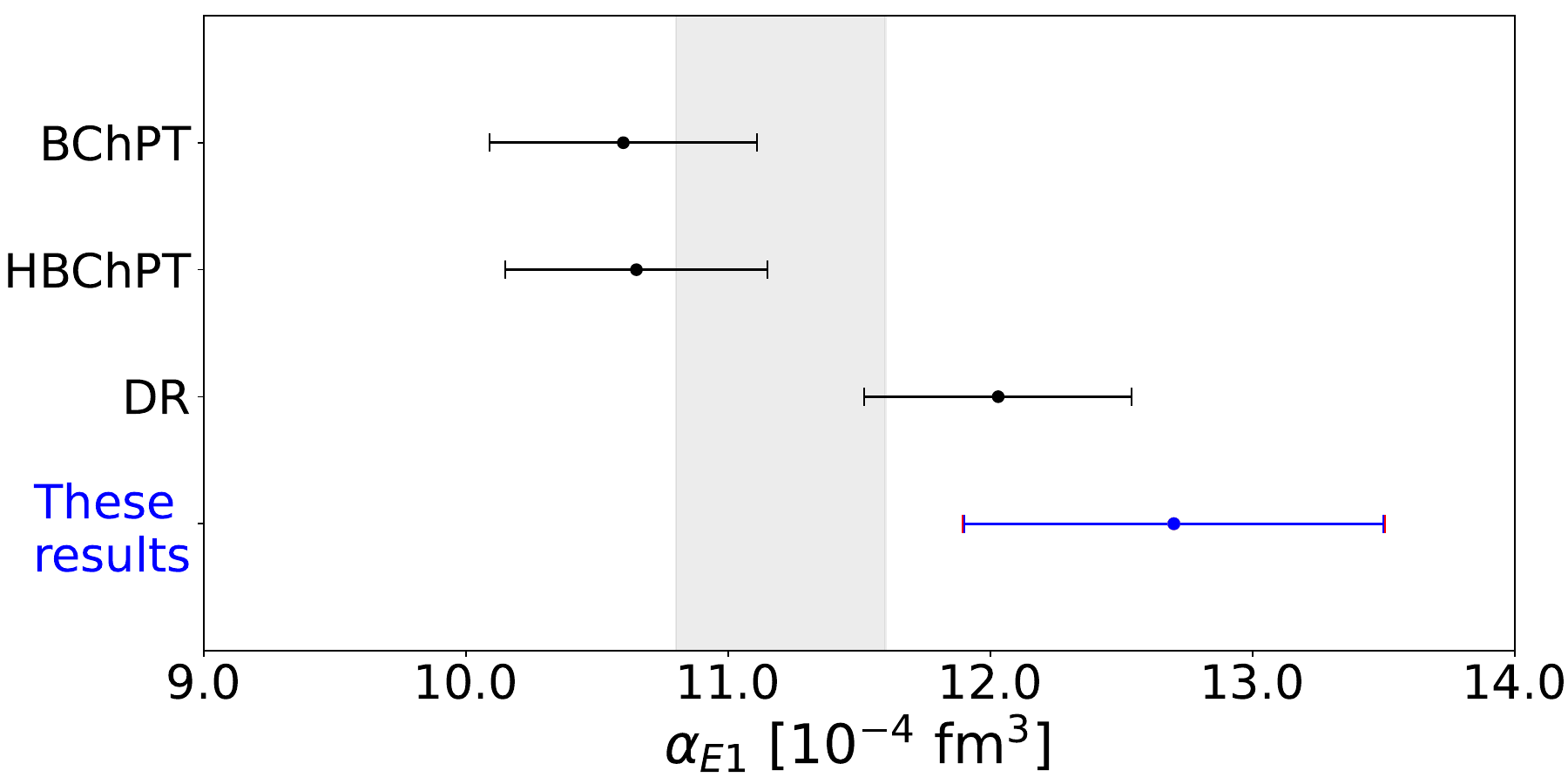}}%
    \subfloat[\label{fig:Results:beta}$\beta_{M1}$]{\includegraphics[width = 0.33\linewidth]{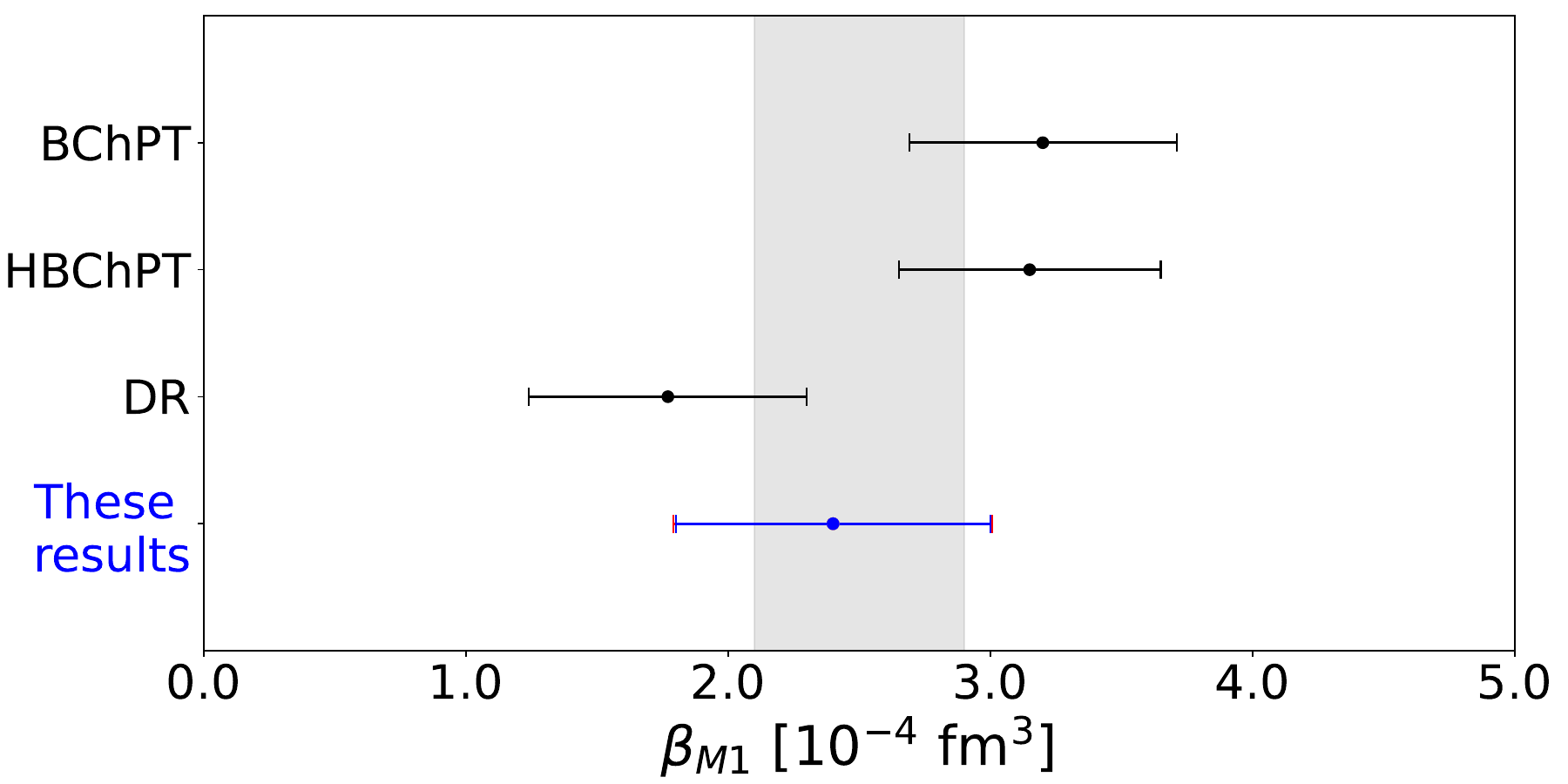}}%
    \subfloat[\label{fig:Results:gE1}$\gamma_{E1E1}$]{\includegraphics[width = 0.33\linewidth]{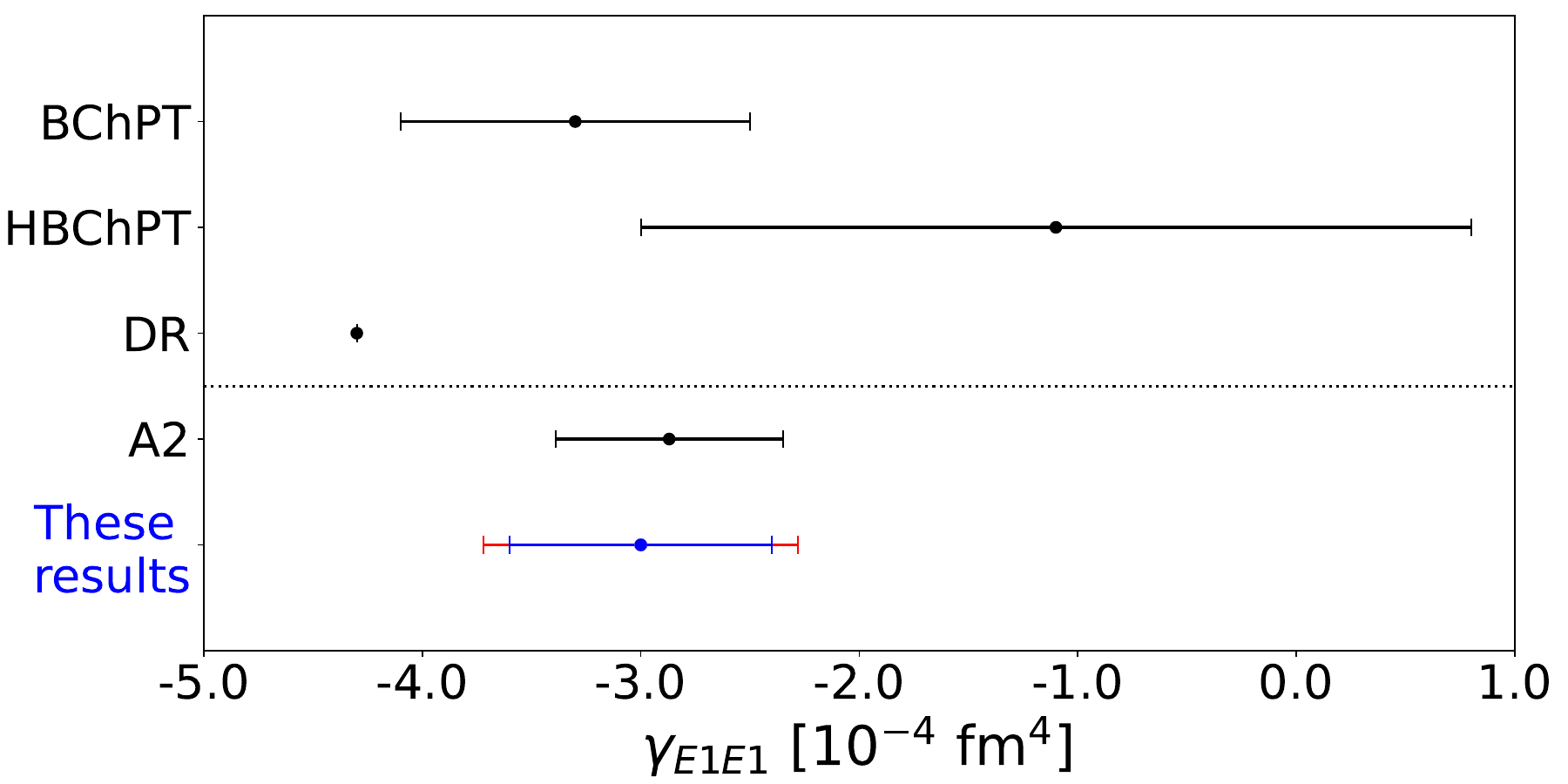}}\\%
    \subfloat[\label{fig:Results:gM1}$\gamma_{M1M1}$]{\includegraphics[width = 0.33\linewidth]{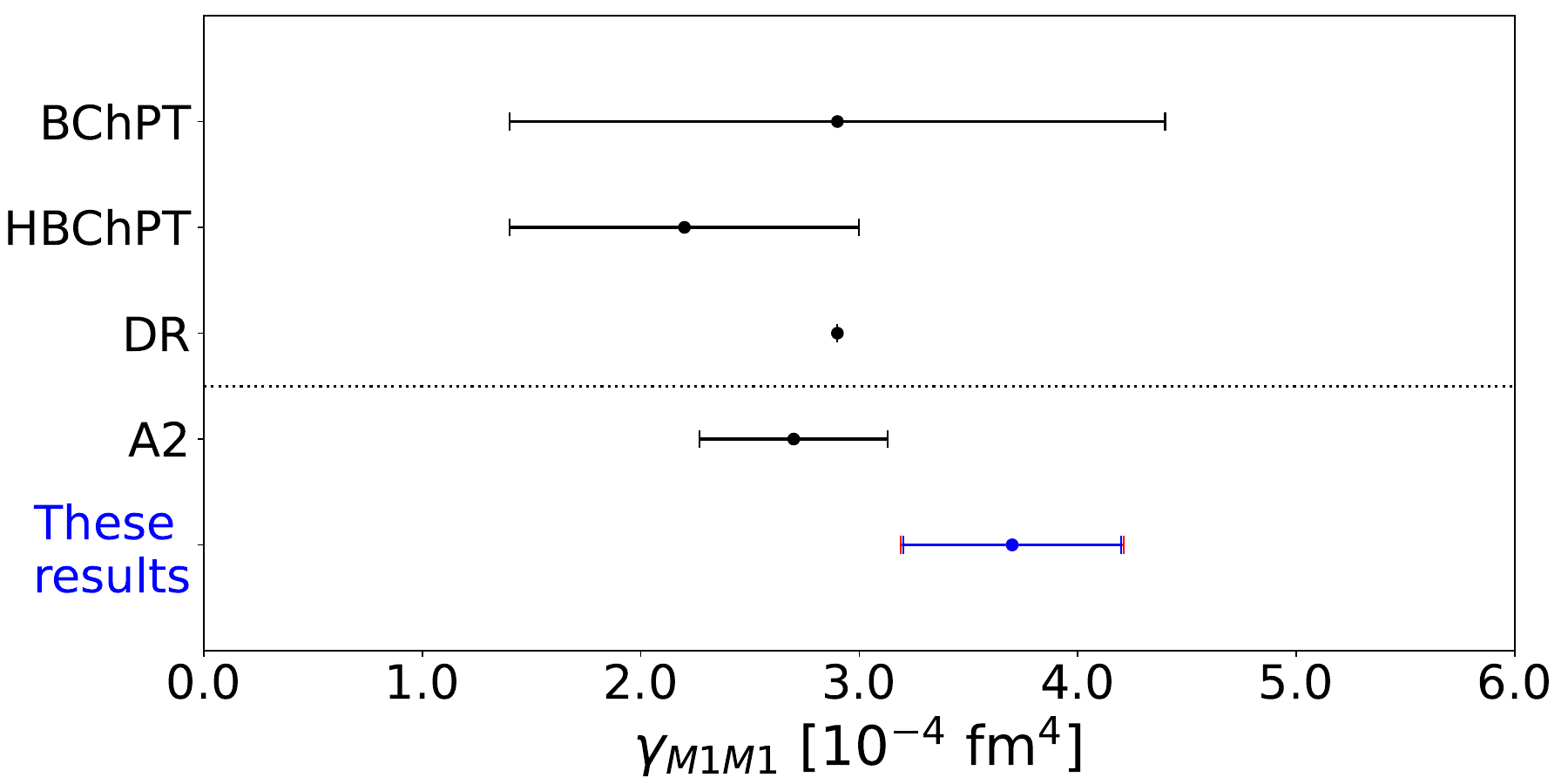}}%
    \subfloat[\label{fig:Results:gM2}$\gamma_{E1M2}$]{\includegraphics[width = 0.33\linewidth]{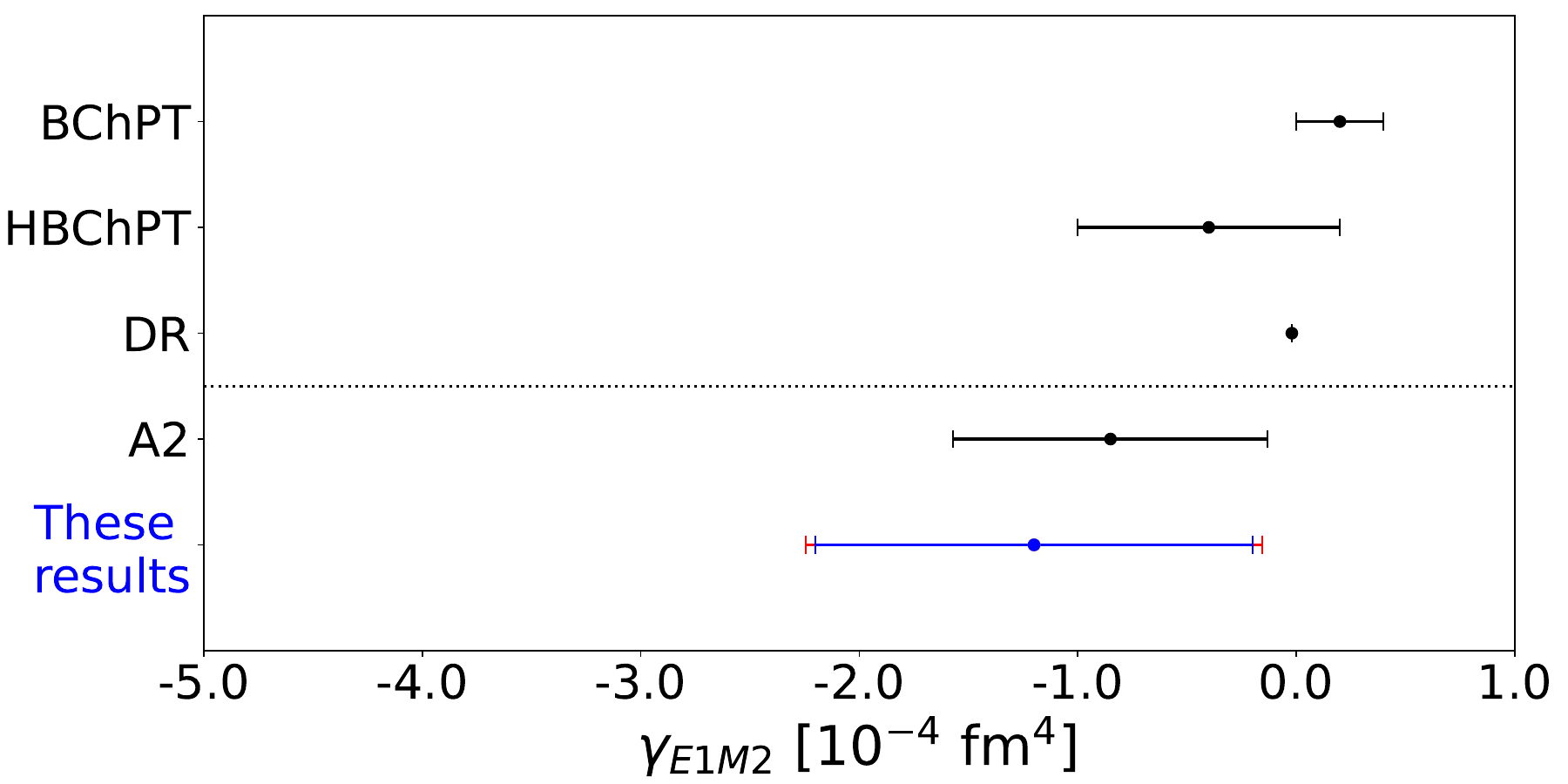}}%
    \subfloat[\label{fig:Results:gE2}$\gamma_{M1E2}$]{\includegraphics[width = 0.33\linewidth]{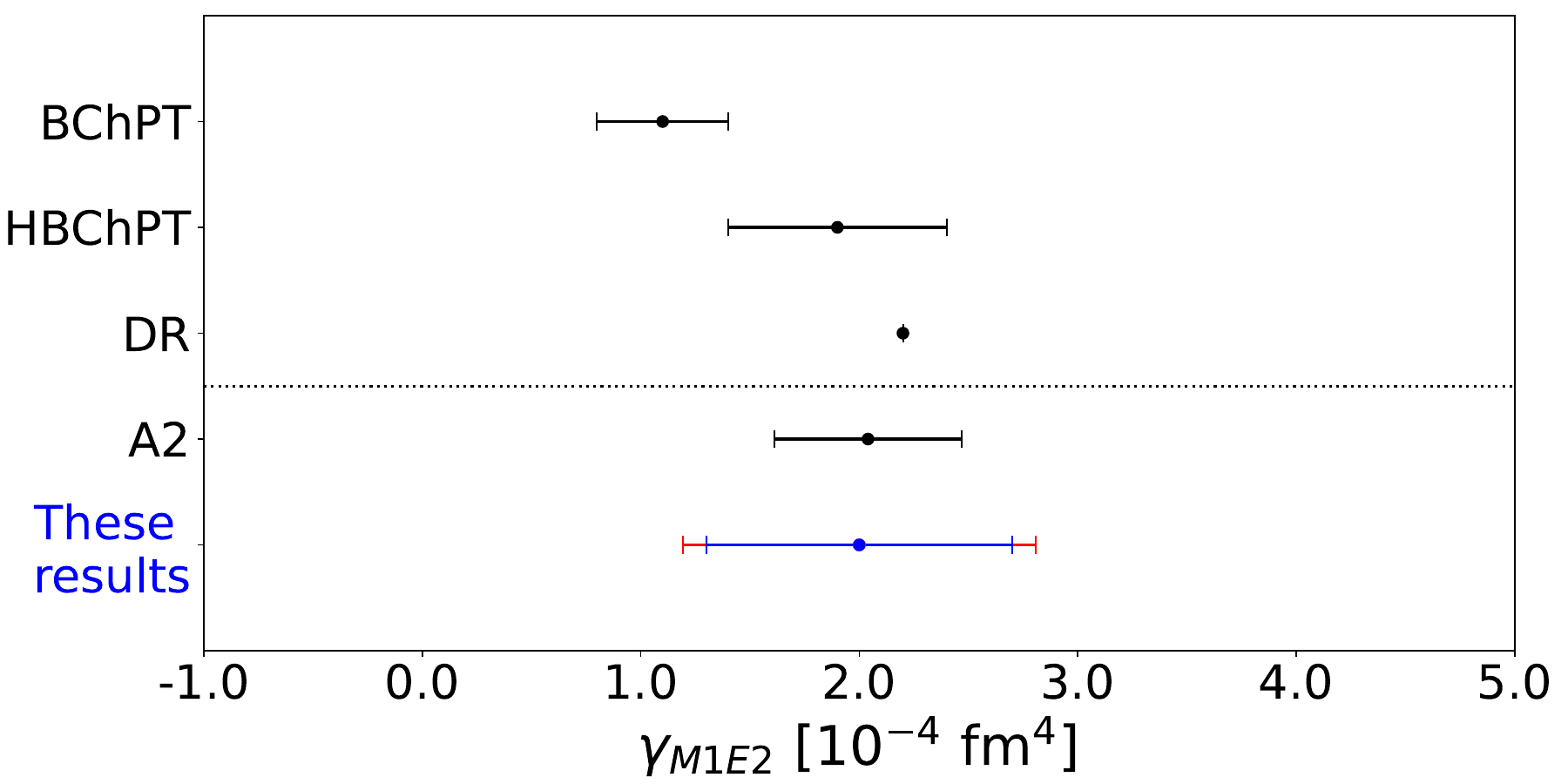}}%
 	\end{center}%
	\caption{Values of the leading-order static proton polarizabilities obtained from the fit procedure. In each panel, our new results are reported in blue. The blue error bars represent the fit error only, the increase in the total error due to the inclusion of the systematic contribution from the model dependency is shown in red. For the two scalar polarizabilities in panels a) and b), DR~\cite{Pasquini:2019nnx}, HBChPT~\cite{McGovern:2012ew}, and BChPT~\cite{Lensky:2014efa} refer to three extractions within DR~\cite{Drechsel:1999rf,Holstein:1999uu,Pasquini:2007hf}, HBChPT~\cite{Griesshammer:2016,Griesshammer:2016}, and BChPT~\cite{Lensky:2009uv} frameworks, respectively. In all cases, most of the spin polarizabilities were fixed to a given value. For the four spin polarizabilities in panels c) to f), some of the existing extractions and theoretical predictions are shown below and above the dotted line, respectively. In particular, A2 refers to the most recent experimental extraction~\cite{A2:2019bqm}, where both scalar polarizabilities were fixed to the PDG values~\cite{ref:PDG}. DR~\cite{Holstein:1999uu}, HBChPT~\cite{McGovern:2012ew,Griesshammer:2016}, and BChPT~\cite{Lensky:2015awa} above the dotted line refer to theoretical predictions using different approaches.}%
	\label{fig:Results}%
\end{figure*}%

A total of $N=10^4$ bootstrapped samples of the database was generated and the minimization was
performed at the end of each iteration with all six proton polarizabilities treated as free
parameters.
For convenience, we used as actual fit parameters some linear combinations of the
scalar and spin polarizabilities: $\alpha_{E1}\pm\beta_{M1}$, $\gamma_{E1E1}$, $\gamma_{M1M1}$,
$\gamma_0 = -\gamma_{E1E1} - \gamma_{M1M1} - \gamma_{E1M2} - \gamma_{M1E2}$,
and $\gamma_{\pi} = -\gamma_{E1E1} + \gamma_{M1M1} - \gamma_{E1M2} + \gamma_{M1E2}$.
The last term, $\gamma_{\pi}$, is the sum of the dispersive contribution
$\gamma_\pi^{{\rm disp}}$, to be fitted to the data, and the pion-pole contribution,
fixed to $\gamma_\pi^{{\rm \pi^0-pole}} = -46.7$~\cite{Schumacher:2005an}.
The choice of the fit parameters allows for a direct comparison, as a consistency check,
of the fit results for $\alpha_{E1}+\beta_{M1}$ and $\gamma_{0}$
to the available experimental predictions of the
Baldin~\cite{deLeon:2001dnx,Gryniuk:2015,Hagel:2016,Strakovsky:2022tvu}
and Gellmann-Goldberger-Thirring (GGT)~\cite{Pasquini:2010zr, Gryniuk:2015, Hagel:2016, Strakovsky:2022tvu}
sum rule values, respectively, obtained using data for the total photoabsorption cross section.
When present, point-to-point systematic errors were added in quadrature to the statistical errors,
while common systematic scale factors were treated as in the previous bootstrap
extractions by Pasquini~\textit{et al.}~\cite{Pasquini:2018,Pasquini:2019nnx},
namely they are assumed to follow a uniform distribution, unless otherwise specified in the
original publication.
Moreover, when multiple systematic sources are given, the final error is the product of all
the generated random uniform variables.

We performed the minimizations by using the nonlinear least-squares fitting routines of the
\texttt{GSL} library~\cite{gls}. As a consistency check of this procedure, both the gradient
and the simplex methods were used as best-fit algorithms and identical results were obtained.
The entire procedure was performed for the three PWAs. The obtained distributions for each parameter
are reported in the Supplemental Material~\cite{suppMat}.
The parameter distributions obtained using the three different PWA inputs have the same shape
and differ only for a shift in their central values.
For this reason, we evaluated the central polarizability values as the mathematical average of the
three different sets of fit values.
Additionally, the largest of the differences between each set of fit values and the average was used to
estimate an additional model error (conservatively considered as a standard deviation) due to dependence on the PWA used as input in the DRs.
The resulting best-fit values are:
\begin{align*}
    &{}\alpha_{E1} + \beta_{M1} = \left(15.1 \pm 0.7\, \hbox{(fit)} \pm 0.1\, \hbox{(model)}\right)\rm\times 10^{-4} fm^{3}, \\
	&{}\alpha_{E1} - \beta_{M1} = \left(10.3 \pm 1.2\, \hbox{(fit)} \pm 0.2\, \hbox{(model)}\right)\rm\times10^{-4} fm^{3}, \\
	&{}\gamma_{E1E1} = \left(-3.0 \pm 0.6\, \hbox{(fit)} \pm 0.4\, \hbox{(model)}\right)\rm\times10^{-4} fm^{4}, \\
	&{}\gamma_{M1M1} = \left(3.7 \pm 0.5\, \hbox{(fit)} \pm 0.1\, \hbox{(model)}\right)\rm\times10^{-4} fm^{4},  \\
	&{}\gamma_{0} = \left(-1.6_{-1.4}^{+1.3}\, \hbox{(fit)} \pm 0.9\, \hbox{(model)}\right)\rm\times10^{-4} fm^{4},\\
    &{}\gamma_{\pi}^{\rm disp} = \left(9.9_{-2.0}^{+1.9}\, \hbox{(fit)} \pm 0.5\, \hbox{(model)}\right)\rm\times10^{-4} fm^{4}. \stepcounter{equation}\tag{\theequation}\label{eq:fitresults}
\end{align*}
The quoted fit errors are the 68\% confidence level (CL) and include the contribution of both
the statistical and systematic uncertainties of the experimental data.
The additional model-dependent systematic uncertainties, evaluated as explained above, are given in rms units, 
and have to be added in quadrature to the previous ones to get the overall values of the estimated systematic uncertainties.

All different fits gave, within rounding errors, a minimum value of the fit function equal to
$\rm\hat{\chi}^2_{red} = 1.13$. As mentioned before, the expected goodness-of-fit distribution
is not given by the $\chi^2$ function, because of the correlations between points of a same
dataset introduced by the systematic uncertainties. This density was then evaluated in the framework of the bootstrap technique~\cite{Pedroni:2019dlg} and a $p$-value = 24\% was estimated from its cumulative distribution.
A plot of this function and the fit correlation matrix are reported in Fig.~3 and Table~3 of the Supplemental Material~\cite{suppMat}, respectively.

The obtained value of both $\alpha_{E1} + \beta_{M1}$ and $\gamma_0$ are in agreement,
within the quoted errors, with the available estimates of the Baldin and GGT sum rule values
listed in Refs.~\cite{Hagel:2016,Strakovsky:2022tvu}.

The final residual distribution, the average $\chi^2$-per-point for each subset, 
and the comparison between the results of the fit with a sample of experimental data from different observables, both below and above pion threshold, are collected in Figs.\@ 3-5 of the Supplemental Material~\cite{suppMat}. 
Taken together, all these results confirm the validity of our overall fit procedure.
 
From the fit results reported in \cref{eq:fitresults}, we obtained the $68\%$ CL intervals of the six
proton static polarizabilities as:
\begin{align*}
    &{}\alpha_{E1} = \left(12.7 \pm 0.8\, \hbox{(fit)} \pm 0.1\, \hbox{(model)}\right)\rm\times 10^{-4} fm^{3}, \\
	&{}\beta_{M1} = \left(2.4 \pm 0.6\, \hbox{(fit)} \pm 0.1\, \hbox{(model)}\right)\rm\times 10^{-4} fm^{3},\\
	&{}\gamma_{E1E1} = \left(-3.0 \pm 0.6\, \hbox{(fit)} \pm 0.4\, \hbox{(model)}\right)\rm\times10^{-4} fm^{4} , \\
	&{}\gamma_{M1M1} = \left(3.7 \pm 0.5\, \hbox{(fit)} \pm 0.1\, \hbox{(model)}\right)\rm\times10^{-4} fm^{4},  \\
	&{}\gamma_{E1M2} = \left(-1.2 \pm 1.0\, \hbox{(fit)} \pm 0.3\, \hbox{(model)}\right)\rm\times10^{-4} fm^{4}, \\
    &{}\gamma_{M1E2} = \left(2.0 \pm 0.7\, \hbox{(fit)} \pm 0.4\, \hbox{(model)}\right)\rm\times10^{-4} fm^{4} ,  \stepcounter{equation}\tag{\theequation}\label{eq:polresults}
\end{align*}
where the meaning of the fit and model errors is the same as in \cref{eq:fitresults}.

The proton polarizability values reported in~\cref{eq:polresults} are shown
as blue points in \cref{fig:Results} (bottom row in each panel).
The blue horizontal bars represent the errors given by the fit procedure.
The increase of the overall systematic uncertainties due to the inclusion of the model errors is shown in red.
In \cref{fig:Results:alpha} and \cref{fig:Results:beta}, the vertical gray bands give
the average $68\%$ CL interval on $\alpha_{E1}$ and $\beta_{M1}$ as evaluated by
the Particle Data Group (PDG)~\cite{ref:PDG}. 
In the same figures, the new results are compared with some of the existing global extractions of $\alpha_{E1}$ and $\beta_{M1}$ 
using DRs~\cite{Pasquini:2019nnx}, HBChPT~\cite{McGovern:2012ew}, and BChPT~\cite{Lensky:2014efa},
respectively, where at least three spin polarizabilities were kept fixed.
In \cref{fig:Results:gE1}, \cref{fig:Results:gM1}, \cref{fig:Results:gE2},
and \cref{fig:Results:gM2} our new results are compared with the last experimental
extraction~\cite{A2:2019bqm}, where the two scalar polarizabilities were fixed,
and three theoretical calculations within DRs~\cite{Holstein:1999uu},
HBChPT~\cite{McGovern:2012ew,Griesshammer:2016}, and ChPT~\cite{Lensky:2014efa}.
The high quality, as well as the importance, of these new results is highlighted
in all the plots.
In fact, they provide a self-consistent extraction of the six leading-order static
proton polarizabilities without any fitting approximation, and with errors that are
competitive with those of all the existing evaluations, that were all performed
by constraining some of the polarizabilities to reduce the uncertainty on the ones of interest.
In particular, the new results from the A2 Collaboration on the unpolarized cross
section~\cite{A2:2021rcs} were fundamental in reducing the correlations between
$\alpha_{E1}-\beta_{M1}$ and $\gamma_{\pi}$, and between $\alpha_{E1}-\beta_{M1}$ and $\gamma_{M1M1}$.

These new results then are a significant benchmark for all future theoretical and experimental
polarizability estimates.
However, the fit parameters have still relevant uncertainties and, as it can be seen from
the values reported in the Supplemental Material~\cite{suppMat}, there is still a slightly
high correlation between $\gamma_{M1M1}$ and $\gamma_{\pi}$
($\rho_{(\gamma_{M1M1}-\gamma_{\pi})} \simeq 0.80$).
These are clear indications of the need for new dedicated measurements,
that will be discussed in detail in a dedicated forthcoming publication.

In summary, we presented the first simultaneous and self-consistent extraction of the six 
leading-order static
proton polarizabilities.
The fit was performed using a bootstrap-based technique combined with a fixed-$t$ subtracted
Dispersion Relation model for the theoretical calculation, using 
three different PWA solutions as input.
The obtained values have an error that is competitive with the existing extractions, that were
all obtained with the inclusion of constraints on some of the fit parameters.
These results provide new important information to our understanding of the internal electromagnetic
proton structure, and should be used as input for further experimental and theoretical extractions.

\begin{acknowledgments}
We are grateful to Igor Strakovsky and Viktor Kashevarov, who provided us with the SAID-MA19 and MAID-2021 multipole values, respectively.
The work of E.~Mornacchi is supported by the Deutsche Forschungsgemeinschaft (DFG, German Research Foundation), through the Collaborative Research Center [The Low-Energy Frontier of the Standard Model, Projektnummer 204404729 - SFB 1044] and by the European Union’s Horizon 2020 research and innovation program under grant agreement number 824093.
S. Rodini acknowledges the support from the DFG grant under the research Unit FOR 2926, ``Next Generation pQCD for Hadron Structure: Preparing for the EIC'', project number 430824754. 
\end{acknowledgments}
%
%
\end{document}